# Rational design of piezoelectric metamaterials with tailored electro-momentum coupling

Zhizhou Zhang[1], Jeong-Ho Lee[1], and Grace X. Gu[1]*

[1]Department of Mechanical Engineering, University of California, Berkeley, CA 94720, USA
*Corresponding author. Email: ggu@berkeley.edu

**Abstract**

Piezoelectric materials have wide sensing and energy transduction applications due to their inherent coupling of mechanical deformation and electric field. Recent discoveries have revealed that asymmetric or heterogeneous microstructures of piezoelectric composites can create an additional coupling of macroscopic momentum to an electric field termed electro-momentum coupling, introducing a new degree of design freedom. In this work, by employing the homogenization scheme, the physical origin of electro-momentum coupling is explored by a high throughput sweep over the microstructure design space of a piezoelectric composite system. This study shows how material constituent properties and geometrical arrangements can affect electro-momentum coupling and be smartly tailored for applications of interest.

**Introduction**

Energy transduction, which involves the conversion of one energy form to another, plays a critical role in a plethora of applications including the detection and generation of electromagnetic and acoustic waves [1,2]. For decades, piezoelectric materials have been commonly used to realize the transduction between electrical and mechanical energy [3,4]. However, current piezoelectric materials possess near homogeneous properties and structures, thus limiting their performance to a specific frequency range or direction of waves [5]. Heterogeneous piezoelectric composites, on the other hand, open an entirely new design space for next-generation transduction materials with the ability to actively manipulate waves. Analogous to the Willis coupling effect, accounting for the coupling of macroscopic momentum to strain [6,7], a recent theoretical discovery in continuum mechanics has revealed that the macroscopic momentum of piezoelectric composites can be coupled with the electric field by patterning piezoelectric materials into an asymmetric or heterogeneous microstructure [8]. This electro-momentum (EM) coupling effect offers a design strategy in which the functionality of piezoelectric metamaterials is tunable by modulation of external stimuli such as electric fields.

To fully take advantage of the tailored metamaterials, theoretical and experimental works have been extensively carried out to explore rational design principles for metamaterials, achieving unprecedented properties and behaviors [9-15]. Furthermore, to derive extraordinary properties manifested in the macroscopic material behavior, resulting from characteristics such as asymmetric microstructures, various analytical schemes have been developed based on homogenization and effective medium theories [16-19]. As a result, constitutive equations between physical fields of the homogenized metamaterials are related to effective (macroscopic) properties that are modeled by coupling parameters. This implies that the newly introduced additional degree of freedom offered by the EM coupling will boost the performance of various devices composed of piezoelectric metamaterials. However, to date, the effect of material and geometrical parameters on the EM coupling has not been well understood. For instance, the relationships between base materials properties such as piezoelectricity and modulus and spatial material distribution (i.e., heterogeneity) and the EM coupling term is not clear in the literature. These fundamental understandings are crucial for the rational design of piezoelectric metamaterials.

This letter focuses on revealing the effects of material and geometrical parameters on the EM coupling coefficient to understand its physical origin and implication. Specifically, our system is described using a one-dimensional, periodic composite with a repeating unit cell consisting of multiple layers as shown in Fig. 1a. We use a digital material design space, which has been widely used to describe the spatial microstructure arrangement of metamaterials [20-24]. In this kind of configuration, materials are discretely allocated in the layers of the unit cell, and material assignment is taken as a design variable. By enumerating all possible permutations, the effect of material heterogeneity and asymmetry on the EM coupling is analyzed via the inconsecutive and irregular order of material distribution in this digital configuration. Harnessing this knowledge is critical for the design of the next generation of piezoelectric metamaterials with tunable properties.

**Methods**

The mechanical and electrical behavior of piezoelectric materials are represented by their governing equations $\nabla \cdot \boldsymbol{\sigma} + \boldsymbol{f} = \dot{\boldsymbol{p}}$ and $\nabla \cdot \boldsymbol{D} = q$, with stress $\boldsymbol{\sigma}$, body force $\boldsymbol{f}$, linear momentum $\boldsymbol{p}$, electric displacement $\boldsymbol{D}$, and free charge density $q$. These governing equations consist of internal fields that have microscopic constitutive relations that can be summarized as:

$$\begin{pmatrix} \boldsymbol{\sigma} \\ \boldsymbol{D} \\ \boldsymbol{p} \end{pmatrix} = \begin{bmatrix} \boldsymbol{C} & \boldsymbol{B}^T & 0 \\ \boldsymbol{B} & -\boldsymbol{A} & 0 \\ 0 & 0 & \rho \end{bmatrix} \begin{pmatrix} \boldsymbol{\varepsilon} \\ \nabla \cdot \phi \\ \dot{\boldsymbol{u}} \end{pmatrix} \tag{1}$$

with displacement $\boldsymbol{u}$, strain $\boldsymbol{\varepsilon}$, and electric potential $\phi$, in which the overdot denotes the time derivative, and the constitutive equations depend on material properties of the stiffness tensor $\boldsymbol{C}$, the dielectric, $\boldsymbol{A}$, and piezoelectric $\boldsymbol{B}$, tensors, and the mass density $\rho$. As revealed by the Willis coupling effect [4,25-30], while the homogenization of complex materials makes the effective (macroscopic) fields identically satisfy the governing equations, the constitutive relations (1) may not be directly utilized for materials with specific microstructures. Therefore, the constitutive relations for the effective internal fields are modified in the form:

$$\begin{pmatrix} \langle \boldsymbol{\sigma} \rangle \\ \langle \boldsymbol{D} \rangle \\ \langle \boldsymbol{p} \rangle \end{pmatrix} = \begin{bmatrix} \widetilde{\boldsymbol{C}} & \widetilde{\boldsymbol{B}}^T & \widetilde{\boldsymbol{S}} \\ \widetilde{\boldsymbol{B}} & -\widetilde{\boldsymbol{A}} & \widetilde{\boldsymbol{W}} \\ \widetilde{\boldsymbol{S}}^\dagger & \widetilde{\boldsymbol{W}}^\dagger & \widetilde{\rho} \end{bmatrix} \begin{pmatrix} \langle \boldsymbol{\varepsilon} \rangle \\ \langle \nabla \cdot \phi \rangle \\ \langle \dot{\boldsymbol{u}} \rangle \end{pmatrix}, \tag{2}$$

which includes coupling coefficients corresponding to the Willis coupling $\widetilde{S}$, and the EM coupling, $\widetilde{W}$, in which $\langle \cdot \rangle$ means the ensemble average, and the overtilde denotes effective material properties that are non-local operators, and $(\cdot)^\dagger$ means the adjoint operator. In this Letter, to obtain the coupling coefficients of piezoelectric metamaterials, we adopt the homogenization scheme based on the effective elastodynamic theory of Willis [31-33], which gives the EM coupling coefficient for a one-dimensional infinite periodic composite under zero free charge ($q = 0$) as

$$\widetilde{W}(\xi) = \frac{s\left\langle \frac{B^2}{A} G^{1D}_{,x} \right\rangle_F \langle G^{1D} \rangle_F^{-1} \langle G^{1D} \rho(x') \rangle_F - s\left\langle \frac{B^2}{A} G^{1D}_{,x} \rho(x') \right\rangle_F}{\left\langle \frac{B}{A} \right\rangle + \left\langle \frac{B}{A} G^{1D}_{,x} \right\rangle_F \langle G^{1D} \rangle_F^{-1} \left\langle G^{1D}_{,x'} \left( C(x') + \frac{B(x')^2}{A(x')} \right) \right\rangle_F - \left\langle \frac{B}{A} G^{1D}_{,xx'} \left( C(x') + \frac{B(x')^2}{A(x')} \right) \right\rangle_F} \tag{3}$$

in spatial and time frequency space with the Fourier, $\xi$, and Laplace, $s = -i\omega$ with time frequency $\frac{\omega}{2\pi}$, transform variables, where the free-space Green's function $G^{1D}$ is employed. Here, the subscript $F$ denotes the Fourier transform. As all tensor quantities become scalar in the one-dimensional problem, the boldface on tensors is omitted here and hereafter.

## Results and Discussion

To investigate the effects of spatial material distribution on the EM coupling coefficient $\widetilde{W}$, we divide a 3 mm periodic unit cell into 8 layers with equal lengths as seen in Fig. 1a. Each microstructure layer is assigned one of the five base materials – lead zirconate titanate (PZT4), barium titanate (BaTiO$_3$), polyvinylidene fluoride (PVDF), aluminum oxide (Al$_2$O$_3$), and polymethyl methacrylate (PMMA) – as listed and indexed in Table. 1. The design space is thus represented as a 1×8 vector $[x_1\ x_2\ \ldots\ x_8]$, where the material index $x_i = 1,2,3,4,5$. For instance, [1 1 3 1 1 1 1 1] denotes that the third layer possesses the material 3 (PVDF) while all other layers are made of the material 1 (PZT4). This yields a total number of $5^8$ possible design candidates considering permutation with repetitions. As an effective dynamic property, $\widetilde{W}$ is a function of both spatial frequency (wavenumber) $\xi$, and time frequency $\frac{\omega}{2\pi}$. Within the long wavelength range ($kl \ll 1$ with the unit cell length $l$, and the elastic wavenumber $k$) where the homogenization scheme is meaningful, $|\widetilde{W}|$ is monotonically increasing in both frequency domains. Taking advantage of this monotonic dependency, the frequency domains are considered to be fixed. As a result, in this Letter, $\xi = 666.7$ m$^{-1}$ and $\frac{\omega}{2\pi} = 0.1$ MHz are used to simplify the design space exploration process.

For each design vector, the homogenized EM coupling coefficient $\widetilde{W}$ is calculated (Fig. 1b). It can be observed that the microstructure configuration can significantly affect the EM coupling, resulting in orders of magnitude difference in $|\widetilde{W}|$. We identify three different points in Fig. 1b that represent a homogeneous microstructure [2 2 2 2 2 2 2 2] with no EM coupling effect, an intermediate design [3 5 1 2 1 1 5 3] showing moderate coupling magnitude, and the maximum design [5 4 1 5 1 4 5 3] achieving the highest $|\widetilde{W}|$ among all the $5^8$ permutations. In fact, although not observable from the plot, any microstructure configurations that share the same spatial sequence as the max design (for instance [5 1 4 5 3 5 4 1]) would possess the exact same $\widetilde{W}$, as they are equivalent to each other under infinite periodicity. Interestingly, the microstructure of the maximum design is perfectly symmetric by considering moving half of the 8th layer to the front. This points to the fact that the real part of $\widetilde{W}$ requires the material heterogeneity while the structural asymmetry is necessary for the phase shift associated with the imaginary part of $\widetilde{W}$ that can be interpreted as its directional dependency, i.e., reciprocity. On the other hand, there are two scenarios causing zero EM coupling: the microstructure itself being homogeneous (only holds for one-dimensional configuration), or the microstructure having no piezoelectricity (material 4 or 5).

Besides $\widetilde{W}$, other effective properties are also investigated including $\widetilde{A}$, $\widetilde{B}$, $\widetilde{C}$, and $\widetilde{S}$ as seen in Eq. 2 to understand potential tradeoffs in coupling terms. The effective properties are sorted in ascending order of $|\widetilde{W}|$ and plotted in Fig. 1c in log scale. The plot is smoothed using a moving average method with an interval window size of 100. The homogenized electric permittivity $\widetilde{A}$ shows a highly similar trend to $\widetilde{W}$ as a high permittivity is required for significant EM coupling. The material designs with small EM coupling effects tend to possess minor Willis coupling $|\widetilde{S}|$, since both coupling effects are dictated by the material heterogeneity and asymmetry. However, their trends do not match perfectly due to different dependencies on the base material properties. On the other hand, the homogenized piezoelectricity $\widetilde{B}$ and elasticity $\widetilde{C}$ are not necessarily small as they have different geometrical and material property dependencies from $\widetilde{W}$. At this stage, no clear correlation among $\widetilde{B}$, $\widetilde{C}$ and $\widetilde{S}$ can be found for the material designs with larger $|\widetilde{W}|$. Nevertheless, future work will attempt to analyze the limits on all these effective properties following the energy and reciprocity constraints as discussed in [3].

As discussed above, the EM coupling is heavily affected by microstructural configurations. Hence, we explore the physical origins (base material properties and geometric parameters) of $\widetilde{W}$ to provide rational design guidelines for piezoelectric metamaterials with large EM coupling. Due to the nonlinearity of the homogenization scheme, it is impractical to obtain any high-level interpretation directly from the theory. Therefore, we instead conduct phenomenological analysis on the entire 8-layer design space where all the $5^8$ design vectors are enumerated. Fig. 2a counts the normalized contribution from each of the base materials, where the contribution from material $m$ is calculated as $\sum_{j=1}^{5^8}\left(\left|\widetilde{W_j}\right| \cdot \sum_{i=1}^{8} \mathbb{1}\left(x_i^j = m\right)\right)$ where $x_i^j$ represents the material assignment for layer $i$ of design vector $j$, and $\mathbb{1}$ represents the indicator function. Results show that material 1 has a high contribution as its high permittivity and piezoelectricity are essential base properties for $\widetilde{W}$. Meanwhile, material 5 also demonstrates remarkable importance despite having the lowest value of $A$ and $B$. It is hypothesized that the existence of such non-piezoelectric materials generates large contrast with respect to the piezoelectric ceramics, especially material 1, and thus greatly contributes to the EM coupling by creating heterogeneity. Fig. 2b plots the statistical correlation between $|\widetilde{W}|$ and the four base material properties listed in Table 1. Here, correlation=0 indicates independence while correlation=1 means perfect linearity. Positive correlations are observed with $A$ and $B$, meaning that base materials with higher permittivity and piezoelectricity are expected to produce stronger EM coupling effects, which validates the high contribution from material 1 as seen in Fig. 2a. On the contrary, $C$ and $\rho$ have negative correlation with $\widetilde{W}$ as strain and velocity are higher in softer, lighter materials.

In addition to the base material constituents, the spatial arrangement can significantly affect the composite's effective constitutive properties. Fig. 2c summarizes how different neighbor layer material combinations affect $|\widetilde{W}|$. When neighbor layers are composed of the same base material, the homogeneity is expected to weaken EM coupling. This can be further validated by the max design where no neighbor layers are identical. However, neighbor combinations (1,1) and (5,5) still make valuable contributions as the base materials 1 and 5 themselves are important for $\widetilde{W}$ (Fig. 2a). As expected, the strong piezoelectricity and large material property contrast in combination (1,5) contributes the most to $\widetilde{W}$. In addition, the small modulus and density in combination (3,5) also shows high importance, matching the observation in Fig. 2b. Material 2 tends to be unfavored by $\widetilde{W}$ due to its high modulus and density although it is a piezoelectric material. Moreover, the relationship between total number of layer changes (i.e., the number of nonidentical neighbors, defined as $\sum_{i=1}^{8} \mathbb{1}(x_i \neq x_{i+1})$) and $\widetilde{W}$ are summarized in Fig. 2d and e. Although the trend is not absolute, having more different neighbor combinations does increase $|\widetilde{W}|$ since it brings more heterogeneity, and at least three different layers are needed to achieve significant EM coupling. We also notice when there are only 2 different layers, the microstructure is inversion symmetric under infinite periodicity, resulting in $\widetilde{W}$ always being real (phase shift = 0 or $\pi$) as stated in [34]. When there are more than 3 different layers, no clear trend can be observed for the phase shift in $\widetilde{W}$ due to its dependence on inversion asymmetry rather than geometric heterogeneity.

The conclusions and hypotheses discussed so far have been based on a pixelated discrete design space with parametrized material assignments. However, a generalized one-dimensional microstructure configuration would involve not only material properties, but also geometric parameters. To investigate how the microstructure geometry affects EM coupling, 2D colormaps of $\widetilde{W}$ are visualized for the 3-layer design configuration as seen in Fig. 3 (at $\xi = 666.7$ m$^{-1}$, $\frac{\omega}{2\pi} = 0.1$ MHz). The total length of a unit cell is constrained to be 3 mm, leaving two independent variables: lengths of the first and second layers, both normalized by the unit cell size. We choose to conduct this analysis on the simplified 3-layer configuration for better visualization and interpretation of the data.

When the 3-layer microstructure consists of 3 different base materials out of the 5 candidate materials in Table 1, there exists a total of 20 different material assignments (10 combinations, each having 2 unique permutations). Among all the 20 configurations, 8 representative colormaps are chosen and plotted in Fig. 3a-h. Interestingly, every microstructure can achieve the complex conjugate $\widetilde{W}$ of its permutated variate by adjusting the layer lengths since their spatial order is simply reversed. For instance, layer lengths of [1.7 0.3 1] mm for material combination [1 2 3] has $\widetilde{W}=0.0719-0.0143i$ Cs/m$^3$, and layer lengths of [1.7 1 0.3] mm for material combination [1 3 2] has $\widetilde{W}=0.0719+0.0143i$ Cs/m$^3$. For material designs with more than 3 layers, changing the sequence of the same combination might cause orders of magnitude difference in $\widetilde{W}$. Moreover, we notice that certain material assignments show much higher upper limits in $\widetilde{W}$ compared with other configurations (Fig. 3a, b, e, f). As a tradeoff, these microstructures are highly sensitive to the layer length (singular hotspot in the colormap), which potentially could add challenges to the manufacturing precision. Fig. 3i plots a case where the microstructure consists of only 2 different base materials. In the case of this [1 3 1] material assignment, $\widetilde{W}$ remains unchanged under a fixed length of layer 2. The reason is that the homogenization scheme treats the first and last layers of any microstructure configurations equivalently as one single layer whenever they share the same base material. Results show that when all the layers share the same material, asymmetry or heterogeneity does not exist, and EM coupling will disappear. Therefore, to obtain a highly EM coupled piezoelectric metamaterial, the material arrangement and geometric parameters must be precisely controlled and smartly tailored for applications of interest.

**Conclusion**

In summary, we reveal how microscopic material and geometric design variables play important roles in the tunability of the EM coupling effect in piezoelectric metamaterials. A well-architectured EM coupling effect brings larger manipulability by external stimuli, implying potential applications of piezoelectric metamaterials such as acoustic sensing and energy harvesting. This work focuses mainly on one-dimensional periodic microstructures, which require at least two base materials to realize heterogeneity and spatial continuity or more base materials for asymmetry. It is envisioned that our study will motivate theoretical and experimental efforts for piezoelectric metamaterials in higher dimensional spaces where a single material with rationally designed internal patterns generates EM coupling for a vast array of applications.

**Conflicts of interest**

There are no conflicts of interest to declare.

**Acknowledgements**

This research was supported by DARPA (Fund Number: W911NF2110363). Additionally, the authors would like to acknowledge Extreme Science and Engineering Discovery Environment (XSEDE) Bridges system, from National Science Foundation (Fund Number: ACI-1548562).

**Table**

Table 1. Properties of the five base materials for the piezoelectric composite from reference [8].

| Index | Material | $A$ (nF/m) | $B$ (C/m$^2$) | $C$ (GPa) | $\rho$ (kg/m$^3$) |
|---|---|---|---|---|---|
| 1 | PZT4 | 5.6 | 15.1 | 115 | 7500 |
| 2 | BaTiO$_3$ | 0.97 | 3.64 | 165 | 6020 |
| 3 | PVDF | 0.067 | -0.027 | 12 | 1780 |
| 4 | Al$_2$O$_3$ | 0.079 | 0 | 300 | 3720 |
| 5 | PMMA | 0.023 | 0 | 3.3 | 1188 |

**Figures**

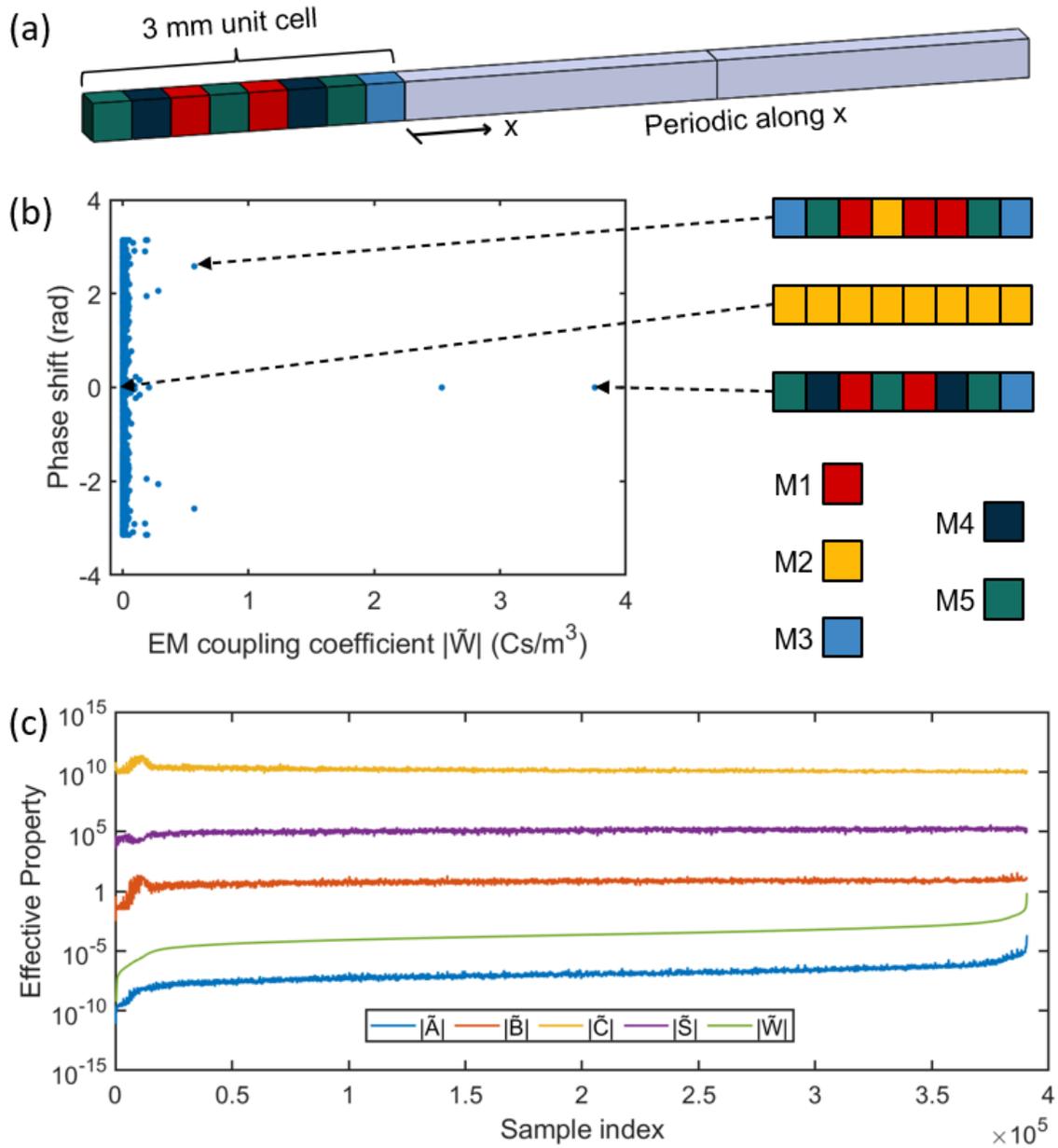

FIG. 1. Complete design sweep for the electro-momentum coupling coefficient of piezoelectric composites. (a) The design space configuration for a 3 mm unit cell consisting of 8 layers with equal lengths. (b) $\tilde{W}$ for all possible permutations with three selected representative configurations of the detailed microstructures. (c) The effective properties (in magnitude) over the entire design space.

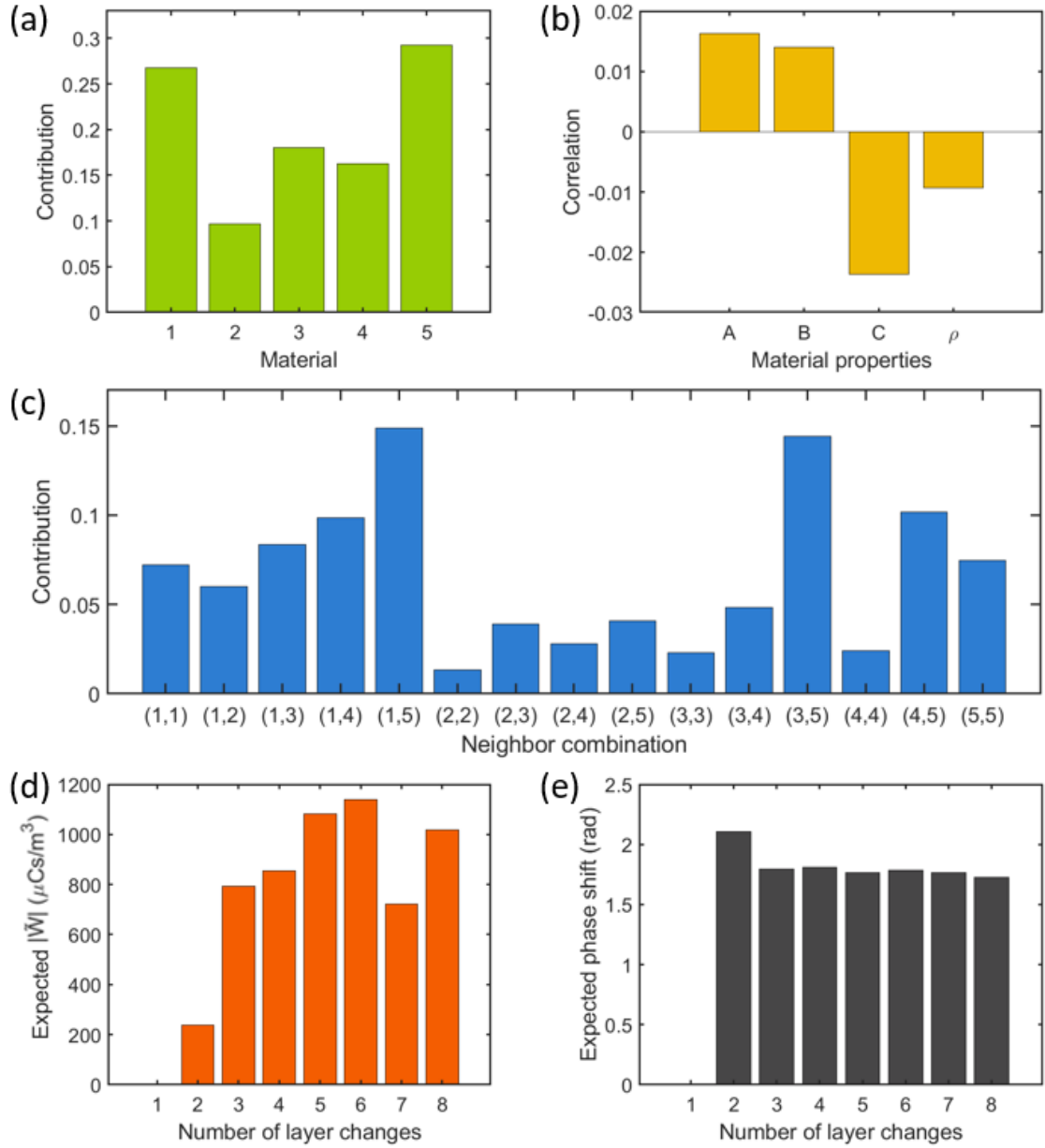

FIG. 2. Data analysis for all the 8-layer microstructure configurations to probe different factors that may affect $\widetilde{W}$. (a) The normalized contributions from different base material selections. (b) The statistical correlation between $|\widetilde{W}|$ and each material property. (c) The effect of different neighbor material combinations on EM coupling magnitude. (d) and (e) The effect of the number of layer changes on $|\widetilde{W}|$ and its phase shift.

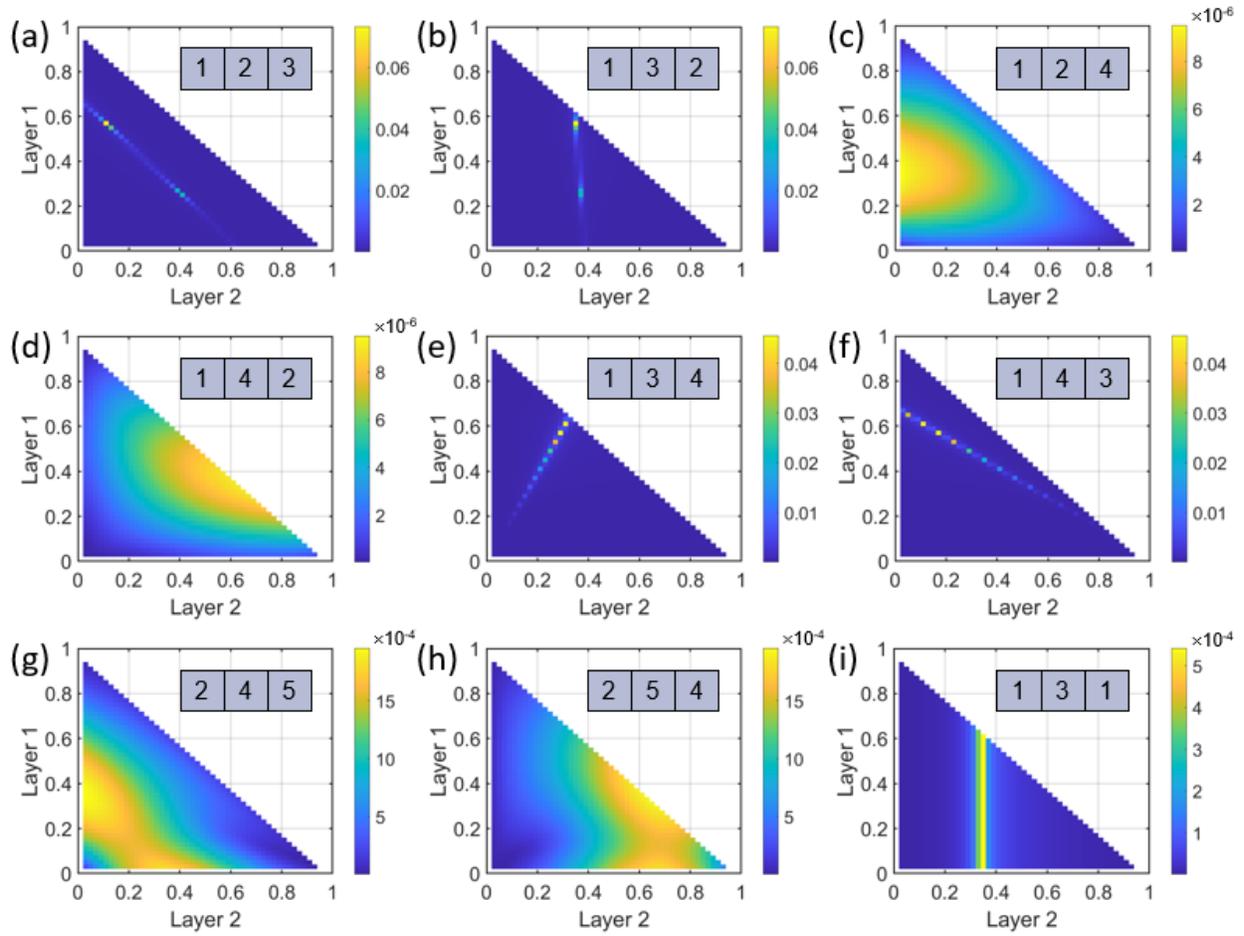

FIG. 3. Visualization of the layer length effect on the magnitude of $\widetilde{W}$ with the colorbar in Cs/m$^3$. Numbers in the boxes indicate the material type of each layer. (a)-(h) Representative colormaps for the unit cell composed of different materials. (i) A representative two-material unit cell in which the middle layer has different material to the others.

10